\documentclass[10pt,letterpaper,twocolumn]{article}   
\usepackage{ol2}
\usepackage{multicol,float}
\usepackage{graphicx}
\pdfoutput=1 

\begin{document}
\twocolumn[
\title{Direct Determination of Diffusion Properties of Random Media from Speckle
Contrast}
\author{Nathan Curry$^{1}$, Pierre Bondareff$^{1}$, Mathieu Leclercq$^{1}$, Niek F. van Hulst$^{2}$, Riccardo Sapienza$^{2}$, Sylvain Gigan$^{1,*}$, Samuel Gr\'{e}sillon$^{1}$}
\address{$^{1}$Institut Langevin, ESPCI, CNRS UMR 7587, UPMC Universit\'{e} Paris 6, 10 rue Vauquelin, 75231 Paris Cedex 05, France.
\\
$^{2}$ICFO-Institut de Ciencies Fotoniques, Mediterranean
Technology Park, 08860 Castelldefels (Barcelona),
Spain, and ICREA-Instituci\'{o} Catalana de Recerca i
Estudis Avan\c{c}ats, 08015 Barcelona, Spain
\\
$^{*}$ Corresponding author: sylvain.gigan@espci.fr
}

\begin{abstract}
We present a simple scheme to determine the diffusion properties of a thin
slab of strongly scattering material by measuring the speckle contrast
resulting from the transmission of a femtosecond pulse with controlled
bandwidth. In contrast with previous methods, our scheme does not require
time measurements nor interferometry. It is well adapted to the
characterization of samples for pulse shaping, non-linear excitation through
scattering media and biological imaging.
\end{abstract}


]


The propagation of optical waves in scattering media is both a fundamental
and a very applied topicwhich requires a deep understanding of how a wave propagates in a
scattering medium\cite{Sebbah2001introduction}. 
An important
transport parameter that governs it is the transport mean
free path, $\ell_t$. 
Upon multiple scattering, that is when the dimension $L$ of the
system is much larger than $\ell_t$,  virtually no ballistic light
transport is present and for $k\,\ell_t\gg 1$ (where $k$ is the
wavenumber) we can treat, in first approximation, the propagation
as a diffusive process governed by the diffusion constant, $D$.

Coherence
  plays an essential role in propagation.
  In the
continuous-wave domain, characterized by a single pulsation
$\omega_0$, multiple scattering gives rise to a very universal
random interference figure: the speckle~\cite{Goodman1976}. For a
polarized monochromatic wave, the spatial distribution of the
speckle intensity $I$ follows
  the
Rayleigh distribution
 $P(I)= \exp (-I/ \langle I \rangle )/\langle I \, \rangle$, where $\langle I \rangle$ is the
mean intensity, and the contrast
 is unity.

In a slab geometry, measuring scattering properties, in particular  $\ell_t$, $D$ or the energy velocity
$v_E$ is by no means trivial.
The angular width of the coherent
backscattering cone~\cite{Wolf1988}, or, in the case of thin slab geometry, 
  total transmission studies~\cite{Genack1987}, give an evaluation of $\ell_t$.
The spread of a
short pulse~\cite{RefShortPulse},  
speckle correlations in
time~\cite{Cai1996} or frequency~\cite{Genack1987}, phase
sensitive interferometric measurements~\cite{Vellekoop2005} or
 variation of the effective refractive index~\cite{Faez2009} allow to extract the diffusion constant $D$.
Simultaneous
measurement of $D$ and $\ell_t$~\cite{RefVelocity,Sapienza2007} give access to $v_E$.

The relevant time, in the dynamical case, is the Thouless
time $\tau_d$\cite{Thouless1977}, with associated bandwidth
$\Delta \nu_d\sim 1/\tau_d$.  The characteristic time
of diffusion  $\tau_d\sim L^2/ (\pi^2 D)$ is the time it takes for a photon to leave the medium, related
 to the
diffuse traversal time $\tau_l\sim L^2/(6 D)$\cite{Landauer1987,Vellekoop2005}, which takes also into account the
direction of the propagation of light ($\tau_l \ge \tau_d$).
In essence, a short laser pulse of duration $\tau_p < \tau_l$
transmitted through the medium will be scattered all over the
medium and  its duration will be extended by approximately
$\tau_l$ by multiple scattering.

Earlier works have studied how the speckle  depends on the
coherence of the source~\cite{Thomson1997} or is modified when a
monochromatic laser is frequency-tuned in the GHz
range~\cite{RefGHz}. The temporal behavior of the speckle has been
measured with a streak camera in the sub-ns range
\cite{Tomita1995} or via non-linear interferometry in the fs range
\cite{WenJun2010}. Some related works on temporal aspects of
speckle from surface scattering have also been studied (see for
instance~\cite{RefTemporalSpeckle}). They rely either on temporal
measurement (direct or via non-linear interferometry or Fourier
Transform) or by time integrated contrast measurement on a camera.

In this paper, we introduce a simple method to determine
the diffusion properties of a thin slab of strongly scattering material, in
a regime where the other techniques are not easily applicable. The method is
based on a simple contrast measurement on a CCD camera and illumination with
a femtosecond laser of variable spectral bandwidth.

The investigated samples are 3D
scattering media with slab geometry made by sedimentation through vertical deposition from a $5$ wt\%
(weight-weight percentage) water suspension of ZnO powder as described in
reference~\cite{Garcia2007}.
This technique allows to grow $\sim$cm$^2$ homogenous samples with
a very flat surface, and filling fraction around 50\%. The
particle size distribution is 230$\pm 70$ nm. 

In the simple model of reference \cite{Goodman1976}, the contrast of the
speckle image is shown to be $C=1/\sqrt{N}$, where  $N$ is the
number of independent speckle patterns added incoherently.
 In the present case, the medium is
illuminated by a pulse with a bandwidth $\Delta\nu_{p}$ larger
than the bandwidth of the medium $\Delta\nu_m \sim
\Delta\nu_d$. After propagation through the medium, there are $N
\simeq \Delta\nu_{p} / \Delta\nu_m$ independent spectral speckles
patterns,
  making the contrast $C\sim
1/\sqrt{\tau_l\times\Delta\nu_p}$. This implies that a measurement
of the contrast as a function of $ \Delta\nu_{p} $ will give
access to $\tau_l$, and therefore $D$.
\begin{figure}[htbp]
\centering
\includegraphics[width=1\linewidth]{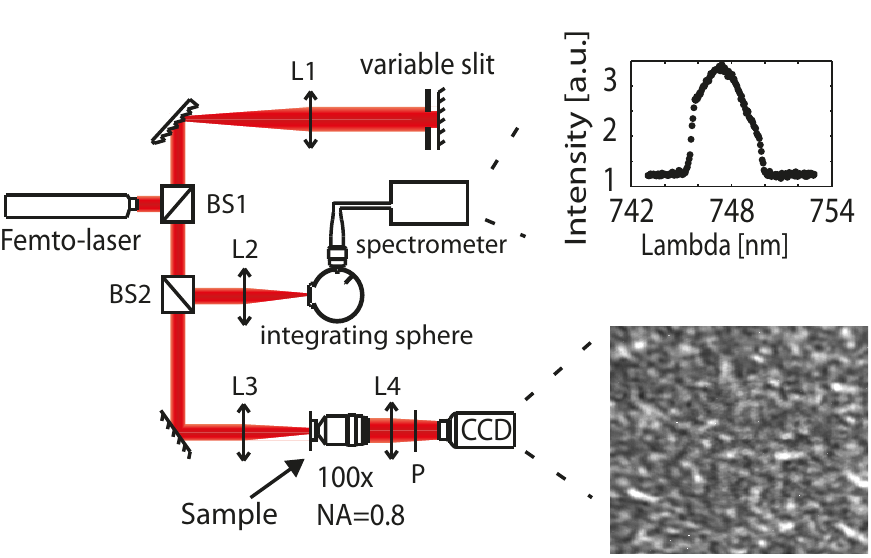}
\caption{Experimental setup with variable slit to control
the spectral bandwidth, typical measured spectrum and speckle image on the
CCD camera}
\label{Figure:Set-up_Contrast}
\end{figure}

The experimental setup is shown in Figure
\ref{Figure:Set-up_Contrast}. The light source is a mode-locked
Ti:Sapphire laser (Spectra Physics Mai Tai 90 fs pulse and 80
MHz repetition rate), with a central wavelength of 748 nm.
 The 2 mm
collimated TEM$_{00}$ output of the laser is sent in a zero
dispersion line~\cite{Monmayrant2010} in order to select the
spectral bandwidth of the light before the scattering medium. It
consists of a series of 3 optical elements, a grating (1200 pitch
per mm), a variable slit attached to a mirror, and a lens between
the mirror and the grating at the focal distance $f = 50\ cm$ from
both elements.
After the zero dispersion line, the beam is focused at the surface
of the scattering medium with a  $f = 50\ cm$ lens. Transmitted
light is collected with a microscope objective (Olympus 100X, NA =
0.8) and the back focal plane of the objective is imaged
with a $f = 20\ cm$ lens on a CCD camera (AVT PIKE F-100B). The
object plane of the microscope is moved by $\sim10$ $ \mu m$ from
the sample surface in order to increase the size of the speckle
grain on the CCD Camera. A linear polarizer is placed between the
objective and the lens to ensure that a single
polarization component is imaged.
To monitor the input spectrum the laser is partially deflected
onto a spectrometer  (Ocean Optics HR4000, resolution
$\le 0.1 nm$) by a beamsplitter after the zero dispersion
line. Spectrum and image are acquired simultaneously. The pulse
bandwidth at the entrance of the medium is between the bandwith of the laser $\Delta
\lambda_{max} = 8\ nm$  and $\Delta \lambda_{min} = 0.3\ nm$, limited
by diffraction on the slit.

A typical image on the CCD is shown in inset on
Fig.~\ref{Figure:Set-up_Contrast}. The contrast of the speckle
is  calculated within a statistically homogeneous region of the
CCD, either in the speckle image (speckle), either not (dark).
The contrast $C=\sigma \textrm{(speckle)}/(\langle
\textrm{speckle} \rangle - \langle \textrm{dark} \rangle ) $ ,
where $\langle \rangle $ and $\sigma\textrm{()} $ are the mean and
standard deviation of the pixel intensities within the defined
region. Error of $C$ is evaluated based on the standard
deviation of the dark image.

Samples with different thicknesses were
illuminated with light pulses of various bandwidths $\Delta \nu_p$ in order to investigate the link between the speckle contrast and $D$. To quantify the contrast modification with  $\Delta \nu_p$ in a more complete formalism,
we treat a thin strongly scattering sample using a slab geometry
and assuming a non-absorbing medium. This treatment, however, is
general and can be extended to any geometry and medium including
those in which both scattering and absorption are significant.
The first question is to determine the distribution of transit
time through such a medium. A general treatment of this problem
has been introduced in~\cite{Patterson1989}. In the multiple scattering
regime,
 the spatially integrated transmittance $T_L (t)$ as a function of time for negligible absorption reads (from Eq. (15) of \cite{Patterson1989}):
\begin{eqnarray}
T_L(t) & = &  (4 \pi D)^{-1/2} \times t^{-3/2} \times \Big\lbrace (L-\ell_t) e^{-\frac{(L-\ell_t)^2}{4 D t}} \label{Equation:Thompson}  \nonumber\\
& & - (L+\ell_t) e^{-\frac{(L+\ell_t)^2}{4 D t}} + (3L-\ell_t) e^{-\frac{(3L-\ell_t)^2}{4 D t}} \nonumber\\
& & +(3L+\ell_t) e^{-\frac{(3L+\ell_t)^2}{4 D t}} \Big\rbrace
.
\end{eqnarray}
We  neglected contributions by more than 3 reflections as they do not contribute significantly to $T_L$ in the numerical calculation. This time-dependent transmission has to be compared to the pulse duration in order to determine the contrast. In reference \cite{Thomson1997} a formalism is introduced  for this purpose, in the context of continuous-wave laser propagating through a very thick medium, which is also valid in the case of a short pulse. A spatially coherent source of spectrum $S(\lambda)$ going through a scattering medium of time distribution $\tilde{T_L}(t)=T(t)/\int_0^{\infty}T(t)dt$ produces a speckle with a contrast $C$ given by the following  integral:
\begin{eqnarray}
C &=& \int_0^{\infty} \frac{1}{S(\lambda)} \lbrace\int_0^{\infty} \int_0^{\infty} S(\lambda) S(\lambda')
\nonumber\\
& &
\times \left\vert f\right(\lambda,\lambda',\tilde{T_L}(t)\left) \right\vert^{2}  d\lambda d\lambda' \rbrace^{1/2}  d\lambda
\label{Equation:ThompsonPatterson}
\end{eqnarray}
and where $f$ is, with a change of variable, a Fourier transform of $\tilde{T_L}(t)$, and reads:
\begin{equation}
f=\int_0^{\infty} c\ \tilde{T_L}(t)\times \exp \left[ -i 2 \pi c\ t (\frac{1}{\lambda}-\frac{1}{\lambda'}) \right]dt .
\end{equation}
It is important to note that all parameters of the problem are not fully independent as $T_L(t)$ and $C$  depend in a non-trivial way on all the parameters $L$, $\ell_t$ and $D$. As a consequence it is not possible to infer all parameters independently from contrast measurements.

\begin{figure}[htbp]
\centering
\includegraphics[width=1\linewidth]{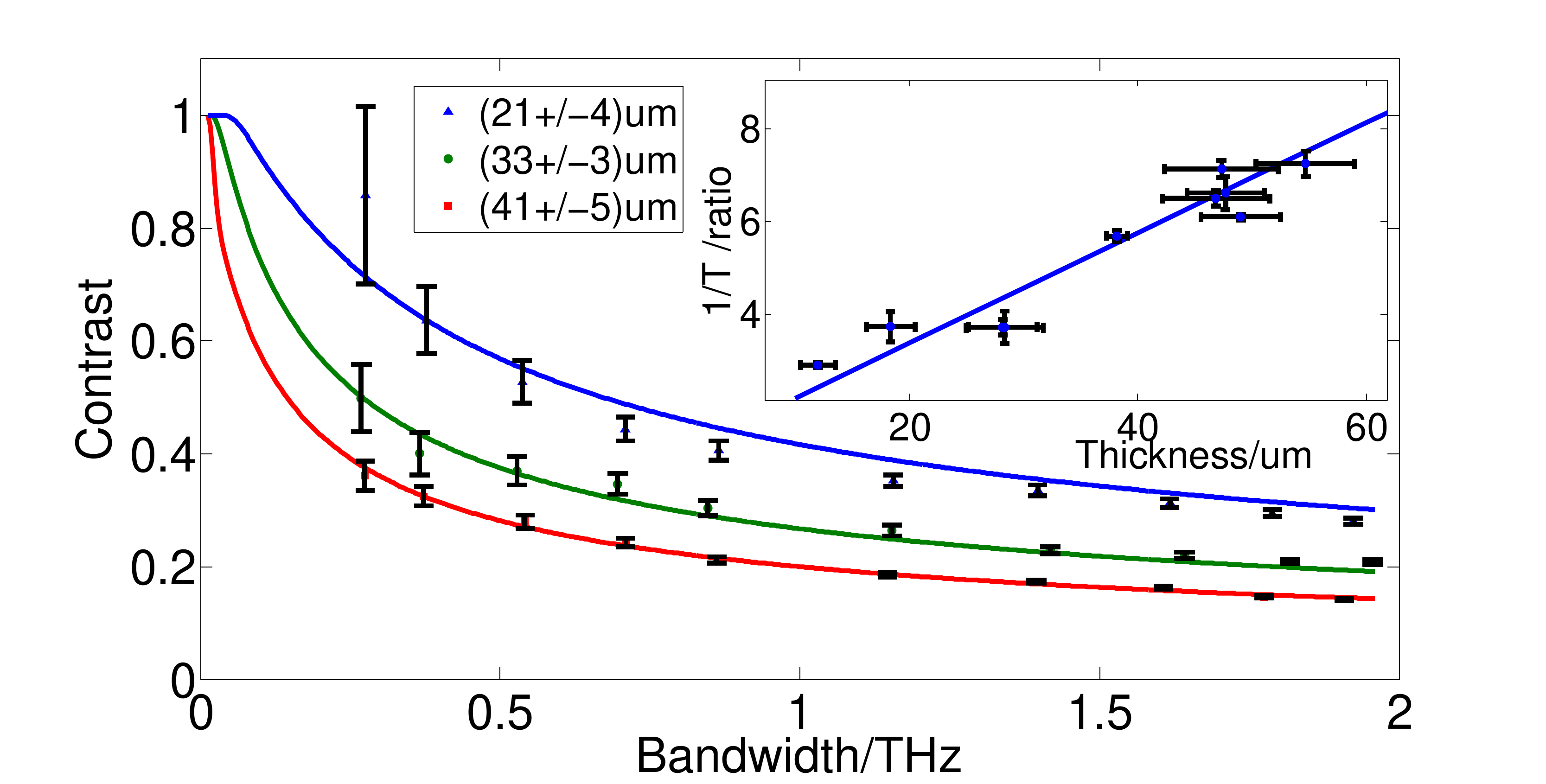}
\caption{Contrast of the speckle for 3 different thicknesses of ZnO. Dots are the experimental measurements, black line is a calculation following the model developed by~\cite{Patterson1989}. Inset: Inverse of the transmission for ZnO layers of different thickness L in $\mu m$.
 }
\label{Figure:Contrast}
\end{figure}

By changing the width of the variable slit, measurements of the
contrast were made over a range of spectral bandwidths
$\Delta\nu_p$. As expected (Fig.~\ref{Figure:Contrast}). As expected
for large values of the bandwidth the contrast asymptotically
decreases with increasing bandwidth in the form $C \sim
1/\sqrt{\Delta\nu_p} $ and decreases with $\Delta\nu_p$ faster for thicker material
as the traveling time gets longer. The monochromatic limit
where $C = 1$ is reached when  $\Delta\nu_p < \Delta\nu_d$, i.e. when $\tau_p$ is identical to the confinement time of the medium~\cite{Cherroret2011}.

Once known $C(\Delta \nu_p)$, the diffusion constant $D$ can be obtained if the transport mean free path $\ell_t$ is measured.
$\ell_t$ can be extracted from a total
transmission measurement by employing the optical analogue of the
Ohm's law: the angular integrated total transmission $T$ is
inversely proportional to the thickness $L$ of the sample. A simple
expression relating $T$ to $L$ is~\cite{Garcia1992}:
\begin{equation}\label{stationary}
T(L) = \frac{\ell_a}{z_e} \frac{\mathrm{sinh}(2 z_e/\ell_a)
\mathrm{sinh}(z_e/\ell_a)}{\mathrm{sinh}((L+2 z_e)/\ell_a)}
\end{equation}
where we have assumed symmetric boundary conditions and 
$z_e =
\frac{\ell_a}{2}\ln(\frac{1+z_0/\ell_a }{1-z_0/\ell_a })$, 
$z_0 = \frac{2\ell_t}{3}\frac{(1+R)}{(1-R)}$,
$\ell_a$ is the absorption length and $R$ is a correction factor
due to the diffuse reflectivity of the boundary
conditions~\cite{Zhu1991}, which we calculated to be $R$=0.49 for
a ZnO sample in air. The interface between the sample and the
glass substrate is here neglected as index matching is almost
achieved between glass (n=1.46) and  the sample
($\langle$n$\rangle\sim$1.5).
Eq. \ref{stationary} holds in the multiple scattering regime ($\ell_t \ll L$) which is the case of the samples under study. When absorption can be neglected, Eq. \ref{stationary} simplifies to
$T(L) \simeq \frac{4}{3}\,\ell_t\,\frac{1+R}{1-R} \, \frac{1}{L}$.

Total transmission as a function of sample thickness is measured
by focusing a laser beam at 748 nm on a set of samples placed at
the entrance of an integrating sphere.  The integrated light
transmitted at all angles is sent via an optical fibre to a
spectrometer, which measures the total transmission
\cite{Sapienza2007}.
The inset in figure~\ref{Figure:Contrast} shows a plot of the inverse
of the transmission $T(L)$ as a function of the thickness for $\lambda=748\ nm$. Vertical error bars are
based on the variation of transmission for different position on
the sample while the horizontal ones are the standard deviation of
the sample thickness upon repeated measurements. Fitting this data
set to the model gave $\ell_t = 2.1
\pm 0.2 \mu m$ and  $\ell_a \ge 100$ $\mu m$
such that absorption is negligible.

$C(\Delta \nu_p)$, $L$ and $\ell_t$ being measured independently, by fitting Eq.
\ref{Equation:ThompsonPatterson} to all data,  $D$  is found to be
$(29\pm4) m^2s^{-1}$ for our ZnO samples.  Fits are shown in Fig.
\ref{Figure:Contrast} for 3 samples. This allows $v_E$ to be
calculated for our medium as $(0.47\pm0.04)\times10^8ms^{-1}$,
i.e. $v_E \simeq 0.16\, c$, where $c$ is the speed of light in
vacuum.

Alternatively, the description of the speckle contrast in terms of
interference of $N$ independent spectral modes provides a
straightforward estimate of the diffusion properties of the medium:
using $C \sim
1/\sqrt{\tau_l\times\Delta\nu_p}$, we estimate the
diffusion coefficient to be $(19\pm2)\ m^2
s^{-1}$, close to the rigorous value previously calculated.

In conclusions, using a CCD camera and a femtosecond light pulse,
we demonstrate a simple set-up to characterize the diffusion
properties of a scattering medium. In order to avoid complex
interferometry measurement of time and phase we exploit the
ability of a multiple scattering medium to mix spatial and
spectral modes at the output~\cite{RefTemporalShaping}.
In view of the recently realized spatio-temporal control of light
pulse in scattering media~\cite{RefTemporalShaping,McCabe2011},
our concept applies for a wide range of Thouless time, and
therefore allows to characterize the diffusion coefficient $D$ for
various scattering media, for photonics and biological
applications.

We acknowledge fruitful discussion with Remi Carminati. PB is funded by ANR ROCOCO. RS and NH acknowledge the financial support of MICINN, programs FIS2009- 08203, CONSOLIDER
CSD2007-046, RyC, Fundacio' CELLEX, and the EU Project ERC.

\end{document}